\documentclass[12pt]{article}
\usepackage{amssymb}
\usepackage{epsfig}
\voffset= - 0.4in
\hoffset= - 0.7in         

\catcode`\@=11
\def\marginnote#1{}

\newcount\hour
\newcount\minute
\newtoks\amorpm
\hour=\time\divide\hour by60
\minute=\time{\multiply\hour by60 \global\advance\minute by-\hour}
\edef\standardtime{{\ifnum\hour<12 \global\amorpm={am}%
        \else\global\amorpm={pm}\advance\hour by-12 \fi
        \ifnum\hour=0 \hour=12 \fi
        \number\hour:\ifnum\minute<10 0\fi\number\minute\the\amorpm}}
\edef\militarytime{\number\hour:\ifnum\minute<10 0\fi\number\minute}

\def\draftlabel#1{{\@bsphack\if@filesw {\let\thepage\relax
   \xdef\@gtempa{\write\@auxout{\string
      \newlabel{#1}{{\@currentlabel}{\thepage}}}}}\@gtempa
   \if@nobreak \ifvmode\nobreak\fi\fi\fi\@esphack}
        \gdef\@eqnlabel{#1}}
\def\@eqnlabel{}
\def\@vacuum{}
\def\draftmarginnote#1{\marginpar{\raggedright\scriptsize\tt#1}}

\def\draft{\oddsidemargin -.5truein
        \def\@oddfoot{\sl preliminary draft \hfil
        \rm\thepage\hfil\sl\today\quad\militarytime}
        \let\@evenfoot\@oddfoot \overfullrule 3pt
        \let\label=\draftlabel
        \let\marginnote=\draftmarginnote
   \def\@eqnnum{(\theequation)\rlap{\kern\marginparsep\tt\@eqnlabel}%
\global\let\@eqnlabel\@vacuum}  }

\def\appname{Appendix}
\newcounter{app}
\def\theapp{\Alph{app}}
\def\app{\par
   \addvspace{4ex}
   \@afterindentfalse
  \secdef\@app\@dapp}
\def\@app[#1]#2{\ifnum \c@secnumdepth >\m@ne
        \refstepcounter{app}
        \addcontentsline{toc}{app}{\theapp
        \hspace{1em}#1}\else
      \addcontentsline{toc}{app}{ #1}\fi
   {\parindent \z@ \raggedright
    \Large \bf \appname~\theapp .
   \Large  \bf 
    #2}\nobreak
   \vskip 4ex   \noindent
\setcounter{equation}{0}
\def\theequation{\Alph{app}.\arabic{equation}}}
\def\@dapp#1{%
{\parindent \z@ \raggedright  \bf #1}\par\nobreak}
\def\l@app#1#2{\addpenalty{\@secpenalty}%
   \addvspace{1em plus\p@}%
   \begingroup
   \@tempdima 3em
     \parindent \z@ \rightskip \@pnumwidth
     \parfillskip -\@pnumwidth
     { \bf
     \leavevmode
     #1\hfil \hbox to\@pnumwidth{\hss #2}}\par
     \nobreak
   \endgroup}

\makeatletter
\newdimen\normalarrayskip            
\newdimen\minarrayskip               
\normalarrayskip\baselineskip
\minarrayskip\jot
\newif\ifold             \oldtrue            \def\new{\oldfalse}
\def\arraymode{\ifold\relax\else\displaystyle\fi}
\def\eqnumphantom{\phantom{(\theequation)}} 
\def\@arrayskip{\ifold\baselineskip\z@\lineskip\z@
     \else
     \baselineskip\minarrayskip\lineskip1\baselineskip\fi}

\def\@arrayclassz{\ifcase \@lastchclass \@acolampacol \or
\@ampacol \or \or \or \@addamp \or
   \@acolampacol \or \@firstampfalse \@acol \fi
\edef\@preamble{\@preamble
  \ifcase \@chnum
     \hfil$\relax\arraymode\@sharp$\hfil
     \or $\relax\arraymode\@sharp$\hfil
     \or \hfil$\relax\arraymode\@sharp$\fi}}

\def\@array[#1]#2{\setbox\@arstrutbox=\hbox{\vrule
     height\arraystretch \ht\strutbox
     depth\arraystretch \dp\strutbox
width\z@}\@mkpream{#2}\edef\@preamble{\halign \noexpand\@halignto
\bgroup \tabskip\z@ \@arstrut \@preamble \tabskip\z@ \cr}%
\let\@startpbox\@@startpbox \let\@endpbox\@@endpbox
  \if #1t\vtop \else \if#1b\vbox \else \vcenter \fi\fi
  \bgroup \let\par\relax
  \let\@sharp##\let\protect\relax
  \@arrayskip\@preamble}
\def\eqnarray{\stepcounter{equation}%
              \let\@currentlabel=\theequation
              \global\@eqnswtrue
              \global\@eqcnt\z@
              \tabskip\@centering              
              \let\\=\@eqncr
              $$%
            \halign to \displaywidth  \bgroup
             \eqnumphantom \@eqnsel
      \hskip\@centering                               
    $\displaystyle  \tabskip\z@ {##}$%
    &\global\@eqcnt\@ne \hskip 2\arraycolsep
         $ \displaystyle  \arraymode{##}$\hfil
    &\global\@eqcnt\tw@ \hskip 2\arraycolsep
         $\displaystyle\tabskip\z@{##}$\hfil
         \tabskip\@centering
    &{##}\tabskip\z@\cr}
\makeatother

\newfont{\hr}{msbm10}
\newfont{\ams}{msam10}

\font\numbers=cmss12
\font\upright=cmu10 scaled\magstep1
\def\stroke{\vrule height8pt width0.4pt depth-0.1pt}
\def\topfleck{\vrule height8pt width0.5pt depth-5.9pt}
\def\botfleck{\vrule height2pt width0.5pt depth0.1pt}
\def\Zmath{\vcenter{\hbox{\numbers\rlap{\rlap{Z}\kern 0.8pt\topfleck}\kern
2.2pt
                   \rlap Z\kern 6pt\botfleck\kern 1pt}}}
\def\Qmath{\vcenter{\hbox{\upright\rlap{\rlap{Q}\kern
                   3.8pt\stroke}\phantom{Q}}}}
\def\Nmath{\vcenter{\hbox{\upright\rlap{I}\kern 1.7pt N}}}
\def\Cmath{\vcenter{\hbox{\upright\rlap{\rlap{C}\kern
                   3.8pt\stroke}\phantom{C}}}}
\def\Rmath{\vcenter{\hbox{\upright\rlap{I}\kern 1.7pt R}}}
\def\Z{\ifmmode\Zmath\else$\Zmath$\fi}
\def\Q{\ifmmode\Qmath\else$\Qmath$\fi}
\def\N{\ifmmode\Nmath\else$\Nmath$\fi}
\def\C{\ifmmode\Cmath\else$\Cmath$\fi}
\def\R{\ifmmode\Rmath\else$\Rmath$\fi}

\def\d{\partial}

\def\bea{\begin{eqnarray}}
\def\eea{\end{eqnarray}}

\def\beq{\begin{equation}}
\def\eeq{\end{equation}}
\def\ba{\beq\new\begin{array}{c}}
\def\ea{\end{array}\eeq}
\def\be{\ba}
\def\ee{\ea}
\def\F{{\cal F}}

\def\stackreb#1#2{\mathrel{\mathop{#2}\limits_{#1}}}

\def\res{{\rm res}}

\def\half{{\textstyle{1\over2}}}
\def\ha{{1\over 2}}
\def\N2{${\cal N}=2$}
\def\4N{${\cal N}=4$}
\def\1N{${\cal N}=1$}
\def\1N*{${\cal N}=1^*$}

\def\beq{\begin{equation}}
\def\eeq{\end{equation}}
\def\ba{\beq\new\begin{array}{c}}
\def\ea{\end{array}\eeq}
\def\be{\ba}
\def\ee{\ea}
\def\theequation{\thesection.\arabic{equation}}
\def\mezzo#1{\bigskip\noindent{\sl #1}\bigskip}

\newcommand{\rf}[1]{(\ref{#1})}

\begin{document}


\begin{flushright}
FIAN/TD-24/07\\
ITEP/TH-57/07
\end{flushright}
\vspace{1.0 cm}

\begin{center}
{\Large\bf
Seiberg-Witten Theory and Extended Toda Hierarchy
}\\
\vspace{1.0 cm}
{\large A.~Marshakov}\\
\vspace{0.6 cm}
{\em
Theory Department, P.N.Lebedev Physics Institute,\\
Institute of Theoretical and Experimental Physics,\\ Moscow, Russia
}\\
\vspace{0.3 cm}
{e-mail:\ \ mars@lpi.ru,\ \ mars@itep.ru}
\end{center}
\begin{quotation}
\noindent
The quasiclassical solution to the extended Toda chain hierarchy, corresponding to the deformation
of the simplest Seiberg-Witten theory by all descendants of the dual topological
string model, is constructed explicitly in terms of the complex curve and generating differential.
The first derivatives of prepotential or quasiclassical tau-function over the extra times,
extending the Toda chain, are expressed through the multiple integrals of the Seiberg-Witten
one-form. We derive the corresponding quasiclassical Virasoro constraints, discuss
the functional formulation of the problem and propose generalization of the extended Toda
hierarchy to the nonabelian theory.
\end{quotation}

\section{Introduction}

The appearance of integrable systems in the context of the Seiberg-Witten theory is now
clearly related to the gauge/string duality. The quasiclassical tau-functions or the
infrared prepotentials, which give the exact low-energy effective actions on the gauge side, become
identified in this framework with
generating functions of the particular topological string models on the string side of duality.
For example, the simplest possible quasiclassical tau-function of extended Seiberg-Witten
theory explicitly
coincides \cite{LMN,MN} with the ``half-truncated'' generating function for the Gromov-Witten
classes on $\mathbb{P}^1$ or the correlation functions of the
topological $\mathbb{P}^1$ string model.

The gauge/string vocabulary looks here as follows: we compare the oversimplified
but perturbed in the ultraviolet, simplest possible ``$U(1)$'' Seiberg-Witten theory
(to be seen, for example, as naive $N_c=1$ particular case of the $U(N_c)$-family) with
the topological string model, describing quantum cohomologies of $\mathbb{P}^1$, to be
generally identified with the base curve of the asymptotically free Seiberg-Witten theory.
The variable $a$, coupled to the unity operator ${\bf 1}$ of string theory, is identified
with the only condensate $\langle\phi\rangle = a$ on the gauge theory side, while the
variable $t_1=\tau_0={\vartheta_0\over 2\pi}+{4\pi i\over g_0^2}$,
coupled to the K\"ahler class $\varpi$ of $\mathbb{P}^1$ target space,
is identified with the (complexified) coupling constant. Moreover, it turns
out, that all perturbations of the gauge theory, encoded in the ultraviolet prepotential
\be
\label{fuv}
F_{UV}(x;{\bf t}) = {\bf t}(x) = \sum_{k>0}t_k{x^{k+1}\over k+1}
\ee
correspond to switching on all gravitational descendants
$\oplus_{k>0}t_{k+1}\sigma_k(\varpi)$ of the K\"ahler class $\varpi$ of the $\mathbb{P}^1$ model,
while the gravitational descendants
of the unity operator remain to be turned off, except for the $\sigma_1({\bf 1})$, which forms the
condensate with $\langle\sigma_1({\bf 1})\rangle\neq 0$. An essential point is that string coupling
$\hbar$ in the $\mathbb{P}^1$ model arises as certain ``equivariant parameter'' of the background,
providing the infrared regularization of the theory on the gauge theory side \cite{Nek}, in
order to collect contributions from the gauge theory instantons, while the instantonic expansion
in gauge theory is going in powers of the scale $\Lambda^2 = e^{{\bf t}''(a)}$.

The exact quasiclassical solution of this theory was explicitly constructed in \cite{MN} as
a solution to dispersionless Toda hierarchy. More generally it was also proposed
for the nonabelian extended Seiberg-Witten theory in terms of quasiclassical tau-function \cite{KriW}
on the deformed by ultraviolet perturbations Seiberg-Witten curve.

However, from the string side of duality this gives rise only to the truncated version of
the $\mathbb{P}^1$ model, and
a natural step would be including the whole set of descendants
$\oplus_{k>0}T_k\sigma_k({\bf 1})$ of another primary - the unity operator.
This has been done already in the $\mathbb{P}^1$ model itself, see \cite{EY,EHY,giv,op,opvir}, where
the matrix integral descriptions was first conjectured, the Virasoro constraints for the
corresponding Gromov-Witten theory were formulated, and the generating functions were constructed
in terms of specific correlators in the theory of free fermions.

Below we are going to write explicitly the quasiclassical solution to this theory,
directly generalizing that of \cite{MN} (see also \cite{AMpopya}). It terms of integrable hierarchies,
it will
raise the dispersionless Toda chain to the so called, following \cite{DubTod,Milan}, extended
Toda hierarchy, where the gravitational descendants $\oplus_{k>0}T_k\sigma_k({\bf 1})$ of unity,
and corresponding ``logarithmic flows'' \cite{EY} extend the set of mutually commuting
flows of the Toda chain.
It turns out, that introducing descendants of unity into the gauge theory is a very
nontrivial step, presumably related to their role of ``deformation'' of the moduli space of
background condensates in field theory, and we will find some hints of that reflected in
the properties of the exact quasiclassical solution.

The extended quasiclassical solution will be constructed in pure
geometric terms, which immediately suggest a natural nonabelian generalization - an extremely
important thing if one would seriously have in mind the application of this duality for the
purposes of gauge theory. The nonabelian generalization is also proposed below, but -
quite typically in the geometric approach - only for class of solutions, when certain finite
number of gravitational descendants of unity is turned on.
We discuss also the relation of our solution to the variational
problem for a certain functional (in spirit of \cite{MN,NO}), in fact even with two
equivalent functional formulations, whose exact relation with the
Nekrasov partition function of
summation over the gauge theory instantons \cite{Nek} remains beyond the scope of this paper.

The paper is organized as follows: in sect.~\ref{sect:disToda} we remind the construction for
the quasiclassical solution to dispersionless Toda chain, corresponding to the half-truncated
$\mathbb{P}^1$ topological string model, with the descendants of unity switched
off, except for a condensate $\langle\sigma_1({\bf 1})\rangle \neq 0$. In sect.~\ref{sect:extToda}
we generalize this solution for the switched on descendants of unity, propose
the formula for the first derivatives of the generating function w.r.t. new variables,
and present explicit
computations for the simplest nontrivial cases of this extension. Next, in sect.~\ref{sect:pert}
we turn first time to the nonabelian theory, and construct the solution corresponding to the
perturbative limit, which produces all important ingredients for the
functional formulation of the problem: the kernel and generalized ultraviolet
prepotential \rf{fuv} for switched on descendants of unity. In sect.~\ref{sect:func} we
discuss the quasiclassical Virasoro constrains and functional formulations of
the problem. Despite the form, suggested by perturbative nonabelian theory, we propose its
equivalent formulation, obtained by an integral transformation and
useful for studying the dependence of the functional upon new times $\{ T_n\}$ of the
extended hierarchy. Finally, in
sect.~\ref{sect:nonabel} we propose the formulation of the nonabelian $U(N_c)$ theory in terms of
abelian differentials on hyperelliptic curve of genus $g=N_c-1$, and discuss the
results and their possible generalizations in sect.~\ref{sect:disc}.

\setcounter{equation}0
\section{Dispersionless Toda chain
\label{sect:disToda}}

Let us, first, remind the main formulas for the solution from \cite{MN} for the dispersionless
Toda chain. We will follow here more convenient normalization from \cite{AMpopya}.

In the case of the deformed in the ultraviolet $U(1)$ supersymmetric
gauge theory the $N_c=1$ Seiberg-Witten curve has a single cut, and the double cover of the
$z$-plane $y^2=(z-x^+)(z-x^-)$ can be always presented in the form
\be\label{u1curve}
z=v+\Lambda\left(w+{1\over w}\right)
\ee
with $x^\pm = v\pm 2\Lambda$ and
\be\label{yu1}
y^2=(z-v)^2-4\Lambda^2
\ee
The solution to dispersionless Toda chain is encoded into the function $S$, odd under
the involution $w\leftrightarrow{1\over w}$
on the double cover \rf{u1curve}, with the asymptotic
\be
\label{sasym}
S(z)\ \stackreb{z\to\infty}{=}\
 - 2z(\log z-1) + \textbf{t}'(z) + 2a\log z - {\d\F\over\d a}
- 2\sum_{k>0} {1\over kz^k}{\d\F\over\d t_k}
\ee
The coefficients at singular terms are identified with the variables of the hierarchy, while
the regular part of expansion defines the first derivatives of the (logarithm of the) tau-function
$\F$. In terms
of the uniformizing variable $w$ one can globally write
\be
\label{Sw}
S = -2z\log w
-2\Lambda(\log\Lambda-1)\left(w-{1\over w}\right)+
\sum_{k>0} t_k\Omega_k(w) + 2a\log w
\ee
where
\be
\label{Ow}
\Omega_k (w) = z^k_+-z^k_-,  \ \ \ k>0
\ee
are the Laurent polynomials, odd under $w\leftrightarrow{1\over w}$. The first
term in \rf{Sw} comes from the Legendre transform of the Seiberg-Witten differential
$d\Sigma \sim z{dw\over w}$.

The canonical Toda chain times are extracted from \rf{sasym} by
\be
\label{t0res}
t_0 = \res_{P_+} dS = - \res_{P_-} dS = 2a
\ee
and
\be
\label{tP}
t_k = {1\over k}\ \res_{P_+} z^{-k}dS =
- {1\over k}\ \res_{P_-} z^{-k}dS,\ \ \ k>0
\ee
From the expansion of $S$
it also immediately follows, that
\be
\label{tPd}
{\d\F\over \d t_k} =  \ha \res_{P_+} z^{k}dS
= - \ha \res_{P_-} z^{k}dS,\ \ \ k>0
\ee
The consistency condition for
\rf{tPd} is ensured by the symmetricity of second derivatives
\be
\label{sysi}
{\d^2\F\over \d t_n\d t_k} = \ha \res_{P_+} (z^k d\Omega_n) = \ha \res_{P_+} (z^n d\Omega_k)
\ee
where
\be
\label{Oz}
\Omega_0 = {\d S\over\d a}\ \stackreb{z\to P_\pm}{=}\ \pm\left(2\log z
- {\d^2 \F\over\d a^2} - 2\sum_{n>0}
{\d^2 \F\over\d a\d t_n}{1\over nz^n}\right)
\\
\Omega_k = {\d S\over\d t_k}\ \stackreb{z\to P_\pm}{=}\ \pm\left(z^k
- {\d^2 \F\over\d a\d t_k} - 2\sum_{n>0}
{\d^2 \F\over\d t_k\d t_n}{1\over nz^n}\right),\ \ \ k>0
\ee
form a basis of meromorphic functions with poles at the points $P_\pm$, with
$z(P_\pm)=\infty$. All time-derivatives here are taken at constant $z$.

Expansion \rf{Oz} of the Hamiltonian functions \rf{Ow} expresses the second
derivatives of $\F$ in terms of the coefficients of the equation of the curve \rf{u1curve}, e.g.
\be
\label{Omexp}
\Omega_0\ \stackreb{z\to\infty}{=}\ 2\log z - 2\log\Lambda - {2v\over z}
- {2\Lambda^2+v^2\over z^2} + \ldots
\\
\Omega_1\ \stackreb{z\to\infty}{=}\ z - v - {2\Lambda^2\over z} - {2v\Lambda^2\over z^2}
+ \ldots
\\
\Omega_2\ \stackreb{z\to\infty}{=}\ z^2 - (v^2+2\Lambda^2) - {4v\Lambda^2\over z} -
{2\Lambda^2(\Lambda^2+2v^2)\over z^2} + \ldots
\ee
Comparison of the coefficients in \rf{Omexp} gives, in particular,
\be
\label{F2der}
{\d^2\F\over \d a^2} = \log\Lambda^2,
\ \ \ \
{\d^2\F\over \d a\d t_1} = v
\ \ \ \
{\d^2 \F \over \d t_1^2} =\Lambda^2
\ee
and, therefore
\be\label{todaeq}
{\d^2 \F \over \d t_1^2} =\exp {\d^2 \F \over \d a^2}
\ee
which becomes the long-wave limit of the Toda chain equations
for the co-ordinate $a^{D} = {{\d\F}\over{\d a}}$
after an extra derivative with respect to $a$ is taken
\be\label{todaequ}
{{{\d}^2 a^D}\over {{\d t_{1}^2}}} \ = \ {\d \over {\d a}}
\exp {\d a^D \over \d a}
\ee
One can now find the dependence of the coefficients of the curve \rf{u1curve}
on the deformation parameters ${\bf t}$ of the microscopic theory by
requiring $dS=0$ at the ramification points $z=x_\pm=v\pm2\Lambda$, where $dz=0$. This condition
avoids from arising of extra singularities at the branch points in the variation of
$dS$ w.r.t. moduli of the curve. Equation
\be
\label{dz}
{dz\over d\log w} = \Lambda\left(w-{1\over w}\right) =0
\ee
fixes the branch points to be at $w = \pm 1$, where now
\be
\label{eqscu}
\left.{dS\over d\log w}\right|_{w =\pm 1} =
\sum_{k>0}t_k \left.{d\Omega_k\over d\log w}\right|_{w =\pm 1}
+ 2a-2v \mp 4\Lambda\log\Lambda =0
\ee
If $t_k=0$ for $k>1$, solution to \rf{eqscu} immediately gives
\be
\label{t01}
v=a, \ \ \ \Lambda^2=e^{t_1}
\ee
and the prepotential
\be
\label{Fsphs}
\F = \ha aa^D + \ha\res_{P_+} \left( z dS\right) -{a^2\over 2} =
\ha a^2t_1+e^{t_1}
\ee
which is a well-known expression for the generating function of the $\mathbb{P}^1$ model,
restricted to the ``small phase space'' of the primary operators.

\bigskip
\mezzo{$\Phi$-function}

\noindent
In the context of dispersionless and generic quasiclassical hierarchies
it is useful to introduce
\be
\label{fiS}
\Phi = {dS\over dz}\ \stackreb{z\to\infty}{=}\
 - 2\log z + \textbf{t}''(z) + {2a\over z} + 2\sum_{k>0} {1\over z^{k+1}}{\d\F\over\d t_k}
\ee
odd under the involution $w \leftrightarrow {1\over w}$, or globally
\be
\label{fiw}
\Phi = -2\log w + \sum_{k>1} kt_k\Omega_{k-1}
\ee
Consistency between \rf{fiw} and \rf{fiS} gives rise exactly to the equations \rf{eqscu}, and
can be used as another way of their derivation. This
function does not have singularities except for the points $P_\pm$ with $z(P_\pm)=\infty$. It
has a natural integral representation
\be
\label{intrepfi}
\Phi(z) = \textbf{t}''(z) - \int dx f''(x)\log(z-x)
\ee
with the integrable ``density'' $f''(x)$
\be
\label{normfpp}
\half\int dx f''(x) = 1,\ \ \ \half\int dx xf''(x) = a
\ee
related to the second derivative of the extremal shape
function for random partitions \cite{NO}. One can easily see, that
\be
\label{fifp}
2if''(z) = \Delta\Phi'(z) = \Phi'(z+i0)-\Phi'(z-i0)
\ee
while for the function \rf{intrepfi} itself one gets
\be
\label{fif}
\Delta\Phi(z) = \Phi(z+i0)-\Phi(z-i0) = - 2i\int dx f''(x)\arg(z-x)  = 2if'(z)
\ee
The function $\Phi$, together with $z$ (or generally one should better refer to their
differentials $d\Phi$ and $dz$ \cite{KriW}), is a basic ingredient for the quasiclassical
hierarchy, and will be exploited below, when discussing the Virasoro constrants.

\setcounter{equation}0
\section{Extended quasiclassical Toda hierarchy
\label{sect:extToda}}

Formula \rf{Sw} can be naturally generalized to the higher logarithmic
flows
\be
\label{SwE}
S = \sum_{k>0}t_k\Omega_k(w) + 2a\log w - 2\sum_{n>0}T_n H_n(z,w)
\ee
so that \rf{Sw} is a particular case of \rf{SwE}, corresponding to
$T_n=\delta_{n,1}$. The extra Hamiltonians
\be
\label{HEk}
H_k(z,w) = z^k\log w + \sum_{j=1}^k C^{(k)}_j\Omega_j(w)
\ee
are odd under involution $w\leftrightarrow{1\over w}$ and fixed by the asymptotic
\be
\label{HEas}
H_k(z,w)\ \stackreb{z\to\infty}{=}\ \pm H_k^{(+)}(z) + O(1)
\\
H_k^{(+)}(z) = z^k(\log z-c_k)
\ee
where the Harmonic numbers
\be
c_k = \sum_{i=1}^k{1\over i},\ \ \ k>0
\ee
(one can also set $c_0=0$)
ensure "scaling property" of the singular parts
\be
\label{scaling}
dH_k^{(+)} = kH_{k-1}^{(+)}dz
\ee
From \rf{HEas} one immediately gets, that
\be
\label{condHE}
C^{(k)}_k = \log\Lambda-c_k = H_k^{(+)}(\Lambda)\Lambda^{-k}
\\
C^{(k)}_j=\omega_{k-j}, \ \ \ j=1, \ldots,k-1
\\
\log w\ \stackreb{z\to\infty}{=}\ \log z - \log\Lambda - \sum_{k>0}{\omega_k\over z^k}
\ee
In particular, $H_0=\log w$, and
\be
\label{H1ey}
H_1(z,w) = z\log w + \Lambda(\log\Lambda-1)\left(w-{1\over w}\right)
\ee
is the Eguchi-Yang term (see \cite{EY}), remaining in the expansion \rf{sasym} for $T_n=\delta_{n,1}$,
which corresponds to nonvanishing condensate $\langle\sigma_1({\bf 1})\rangle\neq 0$.
One can also write for \rf{SwE}
\be
\label{SwEx}
S = -2T(z)\log w + 2a\log w + \sum_{k>0}{\hat t}_k\Omega_k(w)
\ee
with
\be
\label{Tx}
T(z) = \sum_{n>0}T_nz^n
\\
{\hat t}_k = t_k -2T_k(\log\Lambda-c_k)-2\sum_{l>0}\omega_lT_{k+l},\ \ \ \ k>0
\ee
which can be interpreted as a reparameterization $z\rightarrow T(z)$ with certain compensating
transformation of the function \rf{Sw}. The
function $T(z)$, and therefore the times $\{ T_n\}$ can be defined through the jumps of the function
\rf{SwE}, \rf{SwEx}
\be
\label{jumpS}
T(z) = {i\over 4\pi}\Delta S
\ee
or, in a different way, via the residues of derivatives
\be
\label{Tres}
T_n = -{1\over 2n!}\res_{P_+} dS^{(n)} = {1\over 2n!}\res_{P_-} dS^{(n)},\ \ \ \ n\geq 0
\ee
with $S^{(n)} = {d^n S\over dz^n}$.

Now let us propose
the dual to \rf{Tres} formula, which defines the corresponding derivatives of the
prepotential
\be
\label{TdF}
\left.{\d\F\over \d T_n}\right|_{\bf t} = (-)^{n} n!\left(S_n\right)_0
\ee
where
\be
\label{Sn}
{d^nS_n\over dz^n} = S,\ \ \ n\geq 0
\ee
or $S_n$ is the $n$-th primitive of \rf{SwE}, odd under the involution $w\leftrightarrow{1\over w}$
of \rf{u1curve}.
This is a new ingredient in the formulation of quasiclassical hierarchy, going beyond the original
setup of \cite{KriW}. This formula is directly related to the
gravitational dressing of the
primary operators in the (here dual, with the superpotential $z=v+\Lambda\left(w+{1\over w}\right)$ on
the $w$-cylinder) Landau-Ginzburg theory, suggested in \cite{Lossev}.
We propose now, that \rf{TdF}, \rf{Sn} is a strict definition of dependence of the quasiclassical
tau-function upon the times if extended hierarchy, which is trusted by
symmetricity of the corresponding second derivatives of \rf{tPdT} and \rf{TdF}, following from the
Riemann bilinear identities on the cut $w$-cylinder \rf{u1curve}, see Appendix.
The definitions of the prepotential, as a function of Toda chain times ${\bf t}$ remains intact, i.e.
\be
\label{tPdT}
\left.{\d\F\over \d t_k}\right|_{\bf T} =  \ha \res_{P_+} z^{k}dS
= - \ha \res_{P_-} z^{k}dS,\ \ \ k>0
\ee
where the derivatives are now taken at fixed ${\bf T}$.

Instead of \rf{sasym} one can now write for \rf{SwE}
\be
\label{sasymE}
S(z)\ \stackreb{z\to\infty}{=}\
 - 2\sum_{n>0}T_nz^n(\log z-c_n) + \textbf{t}'(z) + 2a\log z - {\d\F\over\d a}
- 2\sum_{k>0} {1\over kz^k}{\d\F\over\d t_k}
\ee
It means, that in addition to \rf{Oz} one gets for the logarithmic Hamiltonians
\be
\label{Hz}
H_n(z,w) = -\ha{\d S(z)\over\d T_n}\ \stackreb{z\to\infty}{=}\ z^n(\log z-c_n) +
\ha {\d^2\F\over\d a\d T_n} + \sum_{k>0} {1\over kz^k}{\d^2\F\over\d T_n\d t_k}
\ee
Note also, that the constant term in the r.h.s. of \rf{TdF} essentially
depend on the negative powers of expansions of $\Omega_k$, therefore ${\d\F\over\d T_n}$ is
expressed in terms of ${\d^2\F\over\d t_k\d t_n}$, and this can be rewritten as a sort of
quasiclassical mixed Hirota-Virasoro type constraints. For example, one gets in this way
\be
\label{dS1exp}
dS_1 = S(z)dz = \sum_{k>0}t_kz^kdz + 2a\log zdz -2\sum_{n>o}T_nH_n^{(+)}(z)dz -{\d\F\over\d a}dz
- 2{\d\F\over\d t_1}{dz\over z} + \ldots
\ee
i.e.
\be
\label{S1exp}
S_1 = \sum_{k>0}{t_k\over k+1}\Omega_{k+1}(w) + 2aH_1(z,w) -
2\sum_{n>o}{T_n\over n+1}H_{n+1}(z,w) -{\d\F\over\d a}\Omega_1(w)
- 2{\d\F\over\d t_1}\log w
\ee
and therefore
\be
\label{S10F}
\left(S_1\right)_0 = -\sum_{k>0}{t_k\over k+1}{\d^2\F\over\d a\d t_{k+1}} + a{\d^2\F\over\d a\d T_1} -
\sum_{n>o}{T_n\over n+1}{\d^2\F\over\d a\d T_{n+1}} +{\d\F\over\d a}{\d^2\F\over\d a\d t_1}
+{\d\F\over\d t_1}{\d^2\F\over\d a^2}
\ee
Upon \rf{TdF} this can be rewritten as
\be
\label{vir?}
{\d\over \d a}\left(a{\d\F\over\d T_1} + {\d\F\over\d a}{\d\F\over\d t_1} -
\sum_{k>0}{1\over k+1}\left(t_k{\d\F\over\d t_{k+1}}+T_k{\d\F\over\d T_{k+1}}\right)\right) = 0
\ee
The quasiclassical Virasoro constraints in their canonical form will be discussed below in
sect.~\ref{sect:func}.

\bigskip
\mezzo{Small phase space}

\noindent
Let now only $t_1$, $a$ and $T_1\neq 1$ be nonvanishing. Then
\be
\label{Sexpz}
S = t_1\Omega_1 + 2a\log w -2T_1 H_1 =
\\
\ \stackreb{z\to\infty}{=}\
t_1z-2T_1z(\log z-1)+2a\log z+\left(2T_1v\log\Lambda-t_1v-2a\log\Lambda\right)-
\\
-\left(
2T_1 \Lambda^2-T_1 v^2-4T_1\Lambda^2\log\Lambda+2T_1 \Lambda^2+2a v\right){1\over z}+
O\left({1\over z^2}\right)
\ee
which means that
\be
\label{S1}
S_1 =
{t_1\over 2}\Omega_2(w)-T_1H_2(z,w)+2aH_1(z,w)+\left(2T_1v\log\Lambda-t_1v-2a\log\Lambda\right)\Omega_1(w)-
\\
-\left(
2T_1 \Lambda^2-T_1 v^2-4T_1\Lambda^2\log\Lambda+2T_1 \Lambda^2+2a v\right)\log w
\ee
and therefore
\be
\label{S10}
\left(S_1\right)_0 =
\ha t_1 v^2-t_1 \Lambda^2-2 T_1 \Lambda^2-2 T_1 v^2\log\Lambda+
4T_1\Lambda^2\log\Lambda+2av\log\Lambda-
\\
-4T_1\Lambda^2 \left(\log\Lambda\right)^2+
2t_1\Lambda^2\log\Lambda
\ee
Equations $\left.{dS\over d\log w}\right|_{w=\pm 1}=0$ now give
\be
\label{vLT}
v = {a\over T_1},\ \ \ \Lambda^2 = \exp{t_1\over T_1}
\ee
which upon substitution into \rf{S10}, and using \rf{TdF} gives rise to
\be
\label{Ft1T1}
\F (t_1,a,T_1) = {a^2t_1\over 2T_1} + T_1^2\exp{t_1\over T_1}
\ee
found originally in \cite{EHY}.
One can conclude therefore, that switching on the first time $T_1$ results
in simultaneous rescaling of all the times $t_1 \to {t_1\over T_1}$, $a \to {a\over t_1}$ etc,
together with the string coupling $\hbar \to {\hbar\over T_1}$, since \rf{Ft1T1} can be
rewritten as
\be
\label{Ft1T1h}
{1\over T_1^2}\F (t_1,a,T_1) = \ha \left({a\over T_1}\right)^2{t_1\over T_1} + \exp{t_1\over T_1} =
\F \left({t_1\over T_1},{a\over T_1};T_1=1\right)
\ee
with the r.h.s. defined in \rf{Fsphs}.

It is interesting to point out that at $T_1\to\infty$, \rf{Ft1T1} gives
\be
\label{Ft1T1inf}
\F (t_1,a,T_1)\ \stackreb{T_1\to\infty}{\sim}\ \left(T_1^2 + T_1t_1 + {t_1^2\over 2}\right) +
{1\over T_1}\left({a^2t_1\over 2} + {t_1^3\over 6}\right) + \ldots =
\\
= \ldots + {1\over 6T_1}\left((t_1+a)^3 + (t_1-a)^3\right) + \ldots =
\ldots {\sf F}(t_1+a,T_1) + {\sf F}(t_1-a,T_1) + \ldots
\ee
modulo quadratic terms and $O\left(T_1^{-2}\right)$, where
\be
\label{topgr}
{\sf F}(x,T_1) = {x^3\over 6T_1}
\ee
is the prepotential of pure two-dimensional topological gravity.

\bigskip
\mezzo{$T_2$ now switched on}

\noindent
Equations \rf{eqscu}
for the switched on $T_2$
(in addition to the small phase space) give rise to
\be
\label{eqT2}
t_1 = (2T_1+4T_2v)\log\Lambda
\\
a = T_1v+T_2\left(v^2 -2\Lambda^2+4\Lambda^2\log\Lambda\right)
\ee
which already cannot be solved analytically for $v$ and $\Lambda$, though the solutions can be easily
found as series in $T_2$, with the first few terms
\be
\label{vLT2}
v = {a\over T_1}-{T_2\over T_1^3} \left(a^2+2t_1T_1e^{t_1\over T_1}-2T_1^2e^{t_1\over T_1}\right)+
{2aT_2^2\over T_1^5} \left(a^2+2t_1^2e^{t_1\over T_1}+2t_1T_1e^{t_1\over T_1}-2T_1^2e^{t_1\over T_1}
\right) + \ldots
\\
\log\Lambda= {t_1\over 2T_1}-{at_1T_2\over T_1^3}+{t_1T_2^2\over T_1^5}
\left(3a^2+2t_1T_1e^{t_1\over T_1}-2T_1^2e^{t_1\over T_1}\right) + \ldots
\ee
and formulas \rf{tPdT}, \rf{TdF} lead to the following expression for the prepotential
\be
\label{FT22}
\F = {a^2t_1\over 2T_1}+T_1^2e^{t_1\over T_1} +T_2\left(
-{a^3t_1\over 3T_1^3}
+4ae^{t_1\over T_1}-{2at_1\over T_1}e^{t_1\over T_1}\right)+
\\
+T_2^2\left({a^4t_1\over 2T_1^5}+{2a^2t_1^2\over T_1^4}e^{t_1\over T_1}
-{2a^2t_1\over T_1^3}e^{t_1\over T_1}
+{t_1^2\over T_1^2}e^{2t_1\over T_1}-{3t_1\over T_1}e^{2t_1\over T_1}
+{5\over 2}e^{2t_1\over T_1}\right) + \ldots
\ee
which certainly satisfies, up to quadratic order in $T_2$, the long-wave limit of
the Toda chain equation \rf{todaeq}.
One can also easily check, that formula \rf{FT22} up to the shift $T_1\to 1-\delta T_1$ from the
condensate of $\langle\sigma_1({\bf 1})\rangle$ and certain rescaling (say, $T_2\to -{T_2\over 2}$)
coincides
with the expansion, obtained in the Appendix of the second paper of \cite{DubTod}.
When deriving \rf{FT22} we have used, in particular, $n=1,2$ cases of \rf{TdF}, expressing
as in \rf{S10} the constant parts of the first two primitives of $S$ (together with
the constant part of the $S$ itself) in terms of the coefficients
of the curve \rf{u1curve}
\be
\label{S10S20}
\left(S\right)_0=2T_1v\log\Lambda-4T_2\Lambda^2+2T_2v^2\log\Lambda+4T_2\Lambda^2\log\Lambda-
vt_1-2a\log\Lambda
\\
\left(S_1\right)_0 =
4T_2v\Lambda^2\log\Lambda-4T_1\Lambda^2(\log\Lambda)^2+
4T_1\Lambda^2\log\Lambda-2T_2v^3\log\Lambda-2T_1v^2\log\Lambda+
\\
+\half t_1v^2-
2\Lambda^2T_1+2av\log\Lambda-8v\Lambda^2T_2(\log\Lambda)^2+
2t_1\Lambda^2\log\Lambda-t_1\Lambda^2
\\
\left(S_2\right)_0 =
T_2v^4\log\Lambda-2a\Lambda^2\log\Lambda-av^2\log\Lambda+
2a\Lambda^2+t_1\Lambda^2v-{1\over 6}t_1v^3+
\\
+T_1v^3\log\Lambda+{5\over 2}T_2\Lambda^4-2T_1v\Lambda^2\log\Lambda-
4T_2v^2\Lambda^2\log\Lambda-6T_2\Lambda^4\log\Lambda+
\\
+4T_2\Lambda^4(\log\Lambda)^2+8T_2v^2\Lambda^2(\log\Lambda)^2+
4T_1v\Lambda^2(\log\Lambda)^2-2t_1v\Lambda^2\log\Lambda
\ee
It is also instructive to write explicitly in this case
\be
\label{fiT2}
\Phi(t_1,a,T_1,T_2) = {dS\over dz} = -2T_1\log w - 4T_2H_1(z,w) =
\\
= -2T_1\log w -4T_2\left(z\log w + \Lambda(\log\Lambda-1)\left(w-{1\over w}\right)\right)
\ee
and
\be
\label{fipT2}
\Phi'(t_1,a,T_1,T_2) = {d\Phi\over dz} = -4T_2\log w -
{2\over\Lambda}{T_1+2T_2v+2T_2\log\Lambda(z-v)\over w-{1\over w}}
\ee
where the coefficients of the curve \rf{u1curve} $\Lambda=\Lambda(t_1,a,T_1,T_2)$
and $v=v(t_1,a,T_1,T_2)$ are constrained by \rf{eqT2}. We see, in particular, that
the Vershik-Kerov ``arcsin law" \cite{VK}, corresponding to the first term in the r.h.s.
of \rf{fiT2} is now not only perturbed by the semicircle Wigner
distribution, (like for the $\sigma_1(\varpi)$ or $t_2$ switched on, see \cite{MN,AMpopya}),
but is also ``modulated''
by multiplication by a linear function. Moreover, one can find, that
\be
\label{fippT2}
\Phi''(t_1,a,T_1,T_2) = {d^2\Phi\over dz^2} =
\\
= {1\over\sqrt{(z-v)^2-4\Lambda^2}}
\left(-4T_2+
{T_1+2T_2v+4T_2\Lambda\log\Lambda\over z-v-2\Lambda}+
{T_1+2T_2v-4T_2\Lambda\log\Lambda\over z-v+2\Lambda}\right)
\ee
For the nonabelian generalization it is also rather useful to rewrite \rf{fipT2} in the form
\be
\label{dfiT2}
d\Phi(t_1,a,T_1,T_2) = -4T_2\log w dz -
4T_2\log\Lambda dy -
2(T_1+2T_2v){dz\over y}
\ee
with $y$ defined in \rf{yu1}.

\bigskip
\mezzo{$T_2,T_3$ switched on}

\noindent
Now, instead of \rf{fiT2}, one gets
\be
\label{fiT3}
\Phi(t_1,a,T_1,T_2,T_3) = {dS\over dz} = -2T_1\log w - 4T_2H_1(z,w) - 6T_3H_2(z,w)
\ee
provided by
\be
\label{eqT3}
t_1 = (2T_1+4T_2v+6T_3v^2)\log\Lambda + 12T_3\Lambda^2(\log\Lambda-1)
\\
a = T_1v+T_2\left(v^2 -2\Lambda^2+4\Lambda^2\log\Lambda\right) +
T_3\left(v^3-6v\Lambda^2+12v\Lambda^2\log\Lambda\right)
\ee
One can easily notice, that in the limit suppressing instantons, i.e. suppressing powers of $\Lambda$
and keeping
only the logarithmic terms $\log\Lambda$, equations \rf{eqT2}, \rf{eqT3} acquire
the form
\be
\label{eqTlam}
t_1 = 2T'(v)\log\Lambda + O(\Lambda^2)
\\
a = T(v) + O(\Lambda^2)
\ee
reflecting the sense of higher descendants of unity as reparameterization $z\to T(z)$.

\setcounter{equation}0
\section{Nonabelian theory: perturbative limit
\label{sect:pert}}

Let us now turn to the problem, how to construct the abelian integral with asymptotic \rf{sasymE} on
generic hyperelliptic curve
\be
\label{dcN}
y^2 = \prod_{j=1}^{2N_c}(z-x_j)
\ee
of the extended nonabelian Seiberg-Witten theory. On the small phase space, i.e. when only the $t_1$ is
nonvanishing, or the descendants of the K\"ahler class are switched off, the curve \rf{dcN} can
be also written as
\be
\label{Todacu}
\Lambda^{N_c}\left(w+{1\over w}\right) = P_{N_c}(z) = \prod_{i=1}^{N_c} (z-v_i)
\ee
with \rf{dcN} turning into
\be
\label{Today}
y^2 = P_{N_c}(z)^2-4\Lambda^{2{N_c}}
\ee
The perturbative limit corresponds to $\Lambda\to 0$ in the above formulas, when the hyperelliptic curve
splits into two disjoint sheets of $z$-plane with $N_c$ punctures, which can be described by
\be
\label{pertcu}
w_{\rm pert} = P_{N_c}(z) = \prod_{i=1}^{N_c} (z-v_i)
\ee
i.e. a rational function on the $z$-plane with $N_c$ punctures. In this section we discuss the perturbative
limit of the nonabelian theory, defined entirely in terms of the rational curve \rf{pertcu}, and turn
to generic situation of \rf{dcN} below in sect.~\ref{sect:nonabel}.

\bigskip
\mezzo{Only $T_2$ switched on}

\noindent
A perturbative anzatz for
\be
\label{dFipert}
\Phi' = -2\sum_{j=1}^{N_c}\left(2T_2\log(z-v_j) + {T'(v_j)\over z-v_j}\right)
\ee
with
\be
\label{CT}
T'(v_j) = T_1 + 2T_2v_j,\  \ \ j=1,\ldots,{N_c}
\ee
can be easily conjectured, having e.g. formula \rf{dfiT2}.
The coefficients of \rf{dFipert} are fixed by
\be
\label{resdFi}
\res_{z=\infty} d\Phi' = 4T_2\cdot N_c
\\
\res_{z=\infty} d\Phi \equiv -\res_{z=\infty} d\Phi' = -2T_1\cdot N_c
\ee
The modified Seiberg-Witten periods are now given by the formulas
\be
\label{aFip}
a_j = {1\over 2\pi i}\oint_{A_j} {z^2\over 2}d\Phi' = \res_{z=v_j} {z^2\over 2}d\Phi' =
\\
= T(v_j) = T_1v_j + T_2v_j^2,\ \ \ \ j=1,\ldots,{N_c}
\ee
Integrating \rf{dFipert} one gets explicitly
\be
\label{Fipert}
\Phi = -2\sum_{j=1}^{N_c}\left(2T_2(z-v_j)(\log(z-v_j)-1) + T'(v_j)\log(z-v_j)\right) +t_1
\ee
One easily finds, that the derivatives of generating differential
\be
\label{derhol}
{\d\over\d a_j}\Phi dz = {1\over T'(v_j)}{\d\Phi\over\d v_j} dz =
2{dz\over z-v_j},\  \ \ j=1,\ldots,{N_c}
\ee
appear to be the ``canonical holomorphic'' differentials with the first order poles
at $z=v_j$ on rational degeneration of the curve \rf{Todacu}. Moreover, one can find, that
\be
\label{derFiT}
-2dH_n = \left.{\d\over\d T_n}\Phi\right|_a dz = \left.{\d\over\d T_n}\Phi\right|_v dz
- \sum_{j=1}^{N_c} v_j^n{dz\over z-v_j}
\ee
giving rise to
\be
\label{dH12}
dH_1 = dz\sum_{j=1}^{N_c}\left(\log(z-v_j)+{v_j\over z-v_j}\right)
\\
dH_2 = dz\sum_{j=1}^{N_c}\left(2z(\log(z-v_j)-1)+2v_j+{v_j^2\over z-v_j}\right)
\ee
where the last terms in the r.h.s. (linear combinations of the ``holomorphic'' differentials
on degenerate rational curve) kill the residue at infinity.

Integrating \rf{Fipert} further, one finds
\be
\label{Spert}
S = t_1z-2\sum_{j=1}^{N_c}\left(T_2(z-v_j)^2\left(\log(z-v_j)-{3\over 2}\right) +
T'(v_j)(z-v_j)(\log(z-v_j)-1)\right) =
\\
= t_1z - 2\sum_{j=1}^{N_c}\left(T_2H_2^{(+)}(z-v_j) + T'(v_j)H_1^{(+)}(z-v_j)\right)
\ee
which defines the perturbative prepotential by
\be
\label{Sprepert}
a^D_i = S(v_i) = t_1v_i- 2\sum_{j\neq i}\left(T_2H_2^{(+)}(v_i-v_j) + T'(v_j)H_1^{(+)}(v_i-v_j)\right)
= {\d\F_{\rm pert}\over\d a_i}
\ee
Formula \rf{Sprepert} can be integrated, since for $i\neq k$ one gets
\be
\label{Sivk}
{\d S(v_i)\over\d v_k} = - 2T'(v_k)\log(v_i-v_k)
\ee
and this gives rise to the perturbative prepotential
\be
\label{pertkern}
\F_{\rm pert} (a_1,\ldots,a_{N_c};T_1,T_2) =
\sum_{j=1}^{N_c} F_{UV}(v_j) + \sum_{i\neq j}F(v_i,v_j;T_1,T_2)
\ee
where one have substitute for $v_i$ a solution to $T(v_i)=a_i$ with the asymptotic
$v_i \sim {a_i\over T_1} + \ldots$, when expanding over the higher times $T_n$. The
bare ultraviolet prepotential
\be
\label{FUV}
F_{UV}(v) = \ha t_1\left(T_1v^2+{4\over 3}T_2v^2\right) =
{a^2t_1\over 2T_1}-{T_2a^3t_1\over 3T_1^3}+{T_2^2a^4t_1\over 2T_1^5}+\ldots
\ee
coincides, of course, with the perturbative part of the $U(1)$ prepotential \rf{FT22} or
partition function of the $\mathbb{P}^1$ model.
The ``interacting part'' in \rf{pertkern} $F(v_i,v_j;T_1,T_2)$
satisfies the integrability condition
\be
\label{prepertaa}
{\d^2 F\over \d a_i\d a_j} = \log(v_i-v_j)
\ee
provided by $a_i = T(v_i)$, $i=1,\ldots,{N_c}$. If only $T_1,T_2\neq 0$, the
direct integration gives an expression
\be
\label{pertkern}
F(v_1,v_2;T_1,T_2) = -\ha (v_1-v_2)^2(T_1+T_2(v_1+v_2))^2\log(v_1-v_2)+
\\
+{1\over 4}(v_1-v_2)^2\left(3(T_1+T_2(v_1+v_2))^2+{1\over 2}T_2^2(v_1-v_2)^2\right) =
\\
= - \ha \left(T(v_1)-T(v_2)\right)^2\log (v_1-v_2)+{3\over 4}\left(T(v_1)-T(v_2)\right)^2
+ {T_2^2\over 8}(v_1-v_2)^4 =
\\
= - \ha \left(a_1-a_2\right)^2\log (v_1-v_2)+{3\over 4}\left(a_1-a_2\right)^2
+ {T_2^2\over 8}(v_1-v_2)^4
\ee
Expanding
over $T_2$ we see that gravitational descendants of unity give rise to the polynomial
corrections to the coupling constants
\be
\label{Tcorr}
T_{ij} = {\d^2\F_{\rm pert}\over\d a_i\d a_j} \sim \log(v_i-v_j) =
\log{a_i-a_j\over T_1} - {T_2\over T_1^2}(a_i+a_j) + O(T_2^2)
\ee
which remind arising in the context of five-dimensional supersymmetric gauge theories. Moreover,
for the particular values $T_n = {(-)^{n-1}\over n}$, we get formally the perturbative
limit of the (compactified) five-dimensional Seiberg-Witten theory \cite{5dSW}, with the infrared
couplings
\be
\label{T5d}
\log(v_i-v_j) = \log\left(e^{a_i}-e^{a_j}\right) = {a_i+a_j\over 2}
+ \log\left(2\sinh{a_i-a_j\over 2}\right)
\ee
and studied recently in the context of its relation to summing over random
partitions in \cite{5dTak}.

\bigskip
\mezzo{Perturbative theory with $N$ descendants of unity switched on}

\noindent
For $N$ first descendants of unity switched on (with arbitrary $N$), it is convenient to introduce
auxiliary functions
\be
\label{fsig}
\sigma(z;x) = \sum_{k>0}{T^{(k)}(x)\over k!}H^{(+)}_k(z-x)
\\
\varphi(z;x) = {d\sigma\over dz} = \sum_{k>1}{T^{(k)}(x)\over (k-1)!}H^{(+)}_{k-1}(z-x)
+ T'(x)\log(z-x)
\ee
with $T^{(k)}(x)$ being $k$-th derivatives of the polynomial $T(x)=\sum_{n=1}^N T_nx^n$.
One can define generally
\be
\label{SFiN}
S(z) = S(z;v_1,\ldots,v_{N_c}) = -2\sum_{j=1}^{N_c} \sigma(z;v_j) + {\bf t}'(z)
\\
\Phi(z) = \Phi(z;v_1,\ldots,v_{N_c}) = -2\sum_{j=1}^{N_c} \varphi(z;v_j) + {\bf t}''(z) = {dS\over dz}
\ee
and express the derivatives of the perturbative prepotential as
\be
\label{Fpsig}
{\d \F_{\rm pert}\over \d v_i} = S(v_i) = {\bf t}'(v_i) - 2\sum_{j\neq i} \sigma(v_i;v_j)
\ee
where the integrability condition \rf{prepertaa} is now ensured by
\be
\label{sigx}
{\d\over \d x}\sigma(z;x) = \sum_{k>0}{T^{(k+1)}(x)\over k!}H^{(+)}_k(z-x)
- \sum_{k>0}{T^{(k)}(x)\over (k-1)!}H^{(+)}_{k-1}(z-x) =
\\
= {T^{(N+1)}(x)\over N!}H^{(+)}_N(z-x) - T'(x)\log(z-x) = - T'(x)\log(z-x)
\ee
We therefore justify formula \rf{pertkern} for arbitrary $N$, i.e.
\be
\label{prepert}
\F_{\rm pert} (a_1,\ldots,a_{N_c};{\bf t},T) = \sum_{j=1}^{N_c} F_{UV}(a_j;{\bf t},T)
+ \sum_{i\neq j}F(a_i,a_j;T)
\\
a_j = T(v_j),\ \ \ \ j=1,\ldots,{N_c}
\\
F_{UV}(a;{\bf t},\textbf{T}) = F_{UV}(v(a);{\bf t},\textbf{T}) =
\int_0^{v(a)} {\bf t}'({\sf v})dT({\sf v})
= \int_0^a {\bf t}'({\sf v}({\sf a}))d{\sf a}
\\
{\d^2\over\d a_i\d a_j}F(a_i,a_j;\textbf{T}) =
\log (v_i(a_i,\textbf{T})-v_j(a_j,\textbf{T}))
\ee
with $v(a)=v(a,T)=T^{-1}(a)$, being a solution with asymptotic $v={a\over T_1} + \ldots$ for
small higher times.

Before considering the nonperturbative formulation on smooth curve \rf{dcN} it is instructive
to discuss the relation of already obtained in $N_c=1$ case formulas with the functional
formulation. As in the half-truncated theory \cite{MN,AMpopya} we postulate, that the linear
and bilinear parts of the functional are directly determined by the perturbative prepotential
\rf{prepert}. In its turn, the functional formulation would become a good ``reference point'' for
the construction in terms of abelian differentials on smooth hyperelliptic curve \rf{dcN}.

\setcounter{equation}0
\section{Functional methods and Virasoro constraints
\label{sect:func}}

Let us now turn to the functional formulation of the proposed above analytic formulas.
To remind, we start first with the case, when all gravitational descendants of unity are
switched off, except for the condensate of $\langle\sigma_1({\bf 1})\rangle \neq 0$.

\bigskip
\mezzo{Switched off $T_n$, $n>1$}

\noindent
The curve \rf{u1curve} endowed with the function \rf{Sw} arises \cite{MN} in the extremum problem
for the functional
\be
\label{functnl}
\F = {1\over 2}\int dx f''(x){\bf t}(x) -
{1\over 2}\int_{x_1>x_2} dx_1 dx_2 f''(x_1)f''(x_2) F(x_1-x_2)
\ee
extremized w.r.t. second derivative of the profile function $f''(x)={d^2 f\over dx^2}$,
constrained by
\be
\label{normlog}
1 = \half\int dx f''(x) = \left.\half f'(x)\right|_{x_-}^{x_+}
\ee
together with
\be
\label{amom}
T_0 = - a = -\half\int dx\ xf''(x) = \half\left.\left(f(x)-xf'(x)\right)\right|_{x_-}^{x_+}
\ee
and where the kernel is
\be
\label{SWkern}
F(x)=\ha H_2^{(+)}(x) = {x^2\over 2}\left(\log x -{3\over 2}\right)
\ee
while the source ${\bf t}(x)$ is defined by ultraviolet prepotential in \rf{fuv}.

Constraints \rf{normlog}, \rf{amom} can be taken into account by adding them to the functional
\rf{functnl} with the Lagrange multipliers
\be
\label{Lagr}
\F \rightarrow \F + a^D\left(a-\half\int dx\ xf''(x)\right) +
\sigma\left(1-\half\int dx\ f''(x)\right)
\ee
so that the variational equation for \rf{Lagr} reads
\be
\label{exeq}
{\bf t}(x) - \int d{\tilde x} f''({\tilde x})F(x-{\tilde x}) = a^Dx + \sigma
\ee
One also gets from \rf{functnl}
\be
\label{derF}
{\d\F\over\d t_k} = {1\over 2(k+1)}\int dx f''(x) x^{k+1},\ \ \ \ k>0
\ee
and, due to \rf{Lagr}
\be
\label{aad}
a^D = {\d\F\over \d a}
\ee
The second Lagrange multiplier in \rf{Lagr}
\be
\label{sig0}
\sigma = -\left(S_1\right)_0 = {\d\F\over \d T_1}
\ee
is given by the derivative of prepotential w.r.t. the first flow of the extended hierarchy.
We remind that the derivatives over the Lagrange multipliers can be taken directly, at constant
$f''(x)$, since all other contributions to these derivatives are proportional to the extremum
equation, and therefore vanish on its solutions.

Integrating \rf{exeq}, one gets the double-integral representation
\be
\label{doubint}
\F = {1\over 2}\int_{x_1>x_2} dx_1 dx_2 f''(x_1)f''(x_2) F(x_1-x_2)+ aa^D + \sigma
\ee
which, together with \rf{functnl}, gives
\be
\label{fhom}
\F = {1\over 2} aa^D + \ha \sigma + {1\over 4}\int dx f''(x)\textbf{t}(x) =
{1\over 2} a{\d\F\over \d a} + \ha \sum_{k>0}t_k {\d\F\over \d t_k} + \ha \sigma
\ee
where the last equality follows from \rf{derF}, \rf{aad}.
Comparing it with representation
\be
\label{Fat}
\F = \ha\left( a{\d\F\over\d a} +
\sum_{k>1}(1-k)t_k {\d\F\over\d t_k}\right) + {\d\F\over\d t_1} -{a^2\over 2}
\ee
and using \rf{sig0}, one derives
\be
\label{L0}
{\d\F\over \d T_1} = - \sum_{k>0}kt_k {\d\F\over \d t_k} + 2{\d\F\over \d t_1} - a^2
\ee
or the quasiclassical $L_0$-Virasoro constraint at fixed $T_n=\delta_{n,1}$.

\bigskip
\mezzo{Quasiclassical Virasoro constraints}

\noindent
The following integral along the boundary of the cut cylinder
\be
\label{virint}
\oint \Phi^2 z^{n+1}dz =
\oint \left({dS\over dz}\right)^2 z^{n+1}dz = 0,\ \ \ n=-1,0,1,2,\ldots
\ee
vanishes, since, analogously to \cite{KriW},
\be
\label{fiexp}
\Phi = {dS\over dz}\ \stackreb{z\to\infty}{=}\
 - 2\sum_{n>0}nT_nz^{n-1}(\log z-c_{n-1}) + \sum_{k>0}kt_k z^{k-1} + {2a\over z}
+ 2\sum_{k=1}^\infty {1\over z^{k+1}}{\d\F\over\d t_k}
\ee
has no singularities in the interior of the cut cylinder, since $dS=0$ at the
branching points, where $dz=0$.

Technically, it is simpler instead of \rf{virint} to consider the ``string equations'', or
the $a$-derivative of this formula. Namely,
\be
\label{seint}
\oint \Phi z^{n+1}{\d\over\d a}\Phi dz = \oint \Phi z^{n+1}{dw\over w} = 0
\ee
since all time-derivatives are taken at constant $z$. Moreover, one can take care only
of the constant part of the contributions into \rf{virint} and \rf{seint} from the $A$- and
$B$- integrals, forming the boundary of the cut cylinder, see details in Appendix.
For example, if $n=-1$ and only $T_1\neq 0$, formula
\rf{seint} gets two obvious contributions
\be
\label{seB}
\left[\int_{B^+}+\int_{B^-}\right]\Phi {dw\over w} \sim T_1\int_B {dw\over w}
\sim T_1{\d^2\F\over\d a^2}
\ee
while
\be
\label{seA}
\left[\int_{A^+}+\int_{A^-}\right]\Phi {dw\over w} \sim
\res_\infty\left(\textbf{t}''(z){dw\over w}\right) \sim t_1 + \sum_{k>1}kt_k{\d^2\F\over\d a\d t_{k-1}}
\ee
which form together the desired string equation, or $a$-derivative of the $L_{-1}$
Virasoro constraint from \cite{EHY,opvir}.

\bigskip
\mezzo{Functional with all descendants switched on}

\noindent
The perturbative formulas in the nonabelian case \rf{prepert} suggest the following form of the functional
with all gravitational descendants stitched on
\be
\label{newfun}
\F = {1\over 2}\int dx f''(x)F_{UV}(x) +
{1\over 2}\int_{x_1>x_2} dx_1 dx_2 f''(x_1)f''(x_2) F(x_1,x_2;\textbf{T}) +
\\
+a^D\left(a-\half\int dxf''(x)T(x)\right) + \sigma\left(1-\half\int dxf''(x)\right)
\ee
where
\footnote{Formulas \rf{newfun} and \rf{cufuv} were derived earlier
by N.~Nekrasov, in a similar context, but using different arguments.}
\be
\label{cufuv}
F_{UV}(x) \equiv F_{UV}(x;{\bf t},\textbf{T}) = \int_0^x {\bf t}'({\sf x})dT({\sf x})
\\
{\d^2\over\d x_1\d x_2}F(x_1,x_2;\textbf{T}) = T'(x_1)T'(x_2)\log(x_1-x_2)
\ee
The variation of \rf{newfun} over $f''(x)$ gives
\be
\label{varnf}
F_{UV}(z) + \int dxf''(x) F(z,x;T)=a^DT(z)+\sigma,\ \ \ z\in {\bf I}
\ee
whose $z$-derivative, after dividing by $T'(z)$, turns into
\be
\label{Seq}
{\bf t}'(z) - \int dxf''(x) \sigma(z;x) = a^D,\ \ \ z\in {\bf I}
\ee
Due to the property of the function $\sigma(z,x)$, following directly from its definition
\rf{fsig} and expansion
\be
\label{Kexp}
{1\over n!}H_{n}^{(+)}(z-x) = {1\over n!}(z-x)^n\left( \log (z-x) - c_{n}\right) =
\\
= \sum_{k=0}^n {H_{n-k}^{(+)}(z)\over (n-k)!} {(-x)^k\over k!} +
(-)^{n-1}\sum_{k>0} {x^{n+k}\over kz^k}{1\over (k+1)\ldots(k+n)}
\ee
one gets (for the switched on $N$ descendants of unity)
\be
\label{sigexp}
\sigma(z;x) = \sum_{k>0}{T^{(k)}(x)\over k!}H^{(+)}_k(z-x) =
\sum_{n=1}^N T_nH^{(+)}_n(z)
- T(x)\log z + \sum_{k>0}{F_k(x)\over kz^k}
\ee
where
\be
\label{fk}
F_k(x) = \int_0^x {\sf x}^k dT({\sf x})
\ee
and we have used the obvious polynomial identities
\be
\label{polid}
k!T_k = T^{(k)}(0)=\sum_{n=0}^N{T^{(n+k)}(x)\over n!}(-x)^n,\ \ \ k=0,\ldots,N
\ee
From \rf{Seq} it follows, that the integral
\be
\label{Ssigma}
S(z) = {\bf t}'(z) - \int dxf''(x) \sigma(z;x) - a^D =
\\
= {\bf t}'(z) - a^D - \sum_{k>0}\int dxf''(x){T^{(k)}(x)\over k!}H^{(+)}_k(z-x)
\ee
whose real part vanishes on the support by \rf{Seq} has an asymptotic
expansion \rf{sasymE} and is constant on the cut. Moreover, the coefficients at negative
powers of $z$ in the r.h.s. are given by
\be
\label{derfunt}
\int dxf''(x) F_k(x) = 2\int dx f''(x){\d F_{UV}(x;{\bf t},\textbf{T})\over \d t_k}
= 2{\d\F\over \d t_k}
\ee
However, it is not easy to get any simple expression for the $T_n$-derivatives of the functional
\rf{newfun}, since almost everything depends on $\{ T_n\}$ in the r.h.s. of this formula. In
order to get the new formula \rf{TdF} for the derivatives over the variables extending the Toda
chain hierarchy,
one has to consider a different form of the functional \rf{newfun}.

\bigskip
\mezzo{Another form of the functional}

\noindent
The formula \rf{Lagr} in fact suggests how the functional problem can be re-formulated in a different way,
when the higher times of extended hierarchy are switched of. Suppose again
that only $T_1,\ldots,T_N$ are
non-vanishing, which somehow characterize the $N$-th ``class of backgrounds" for the gauge theory.
One can write for the perturbative prepotential
\be
\label{fuvD}
F_{UV}(x) = \int_0^x \textbf{t}'({\sf x}) dT({\sf x}) = \textbf{t}(x)T'(x) - \textbf{t}_2(x)T''(x)
+ \ldots + (-)^{N-1}\textbf{t}_N(x)T^{(N)} =
\\
= {\hat D}_{N-1}(x){\bf t}_N(x)
\ee
where
\be
\label{tN}
{\bf t}_N(x) = \sum_{k>0} t_k {x^{k+N}\over (k+1)\ldots(k+N)}
\\
{\bf t}(x) \equiv {\bf t}_1(x) = {d^{N-1}\over dx^{N-1}}{\bf t}_N(x)
\ee
and we have introduced the differential operator with the polynomial coefficients
\be
\label{DN}
{\hat D}_{N-1}(x) = T'(x){d^{N-1}\over dx^{N-1}}-T''(x){d^{N-2}\over dx^{N-2}}+\ldots
+(-)^{N-1}T^{(N)}
\ee
Consider also an integral transform, or
introduce new ``density'' by the formula
\be
\label{rhof}
\int dx\rho(x)g(x) = \int dx f''(x){\hat D}_{N-1}(x)g(x)
\ee
for an integral over the support ${\bf I}$ with an arbitrary function $g(x)$ (from some reasonable
class of functions). It means, that in certain sense this density is
$\rho(x) \sim {\hat D}^\dagger_{N-1}(x)f''(x)$.
Note also, that using ${\hat D}$-operators \rf{fuvD}, one can write for the kernel in \rf{cufuv}
\be
\label{kerFD}
F(z,x) = {(-)^{N}\over (2N)!}{\hat D}_{N-1}(z){\hat D}_{N-1}(x)H_{2N}^{(+)}(z-x) =
\\
= \sum_{n,k=1}^N (-)^{n-1}T^{(n)}(z)T^{(k)}(x){H_{n+k}^{(+)}(z-x)\over (n+k)!}
\ee
while the contribution of the linear term in \rf{newfun} - with the ultraviolet prepotential -
turns into
\be
\int dxf''(x) F_{UV}(x) = \int dx\rho(x){\bf t}_N(x)
\ee
The density $\rho(x)$ obeys important constraints, directly following from \rf{rhof}, namely
\be
\label{rhocstr}
\ha\int dx \rho(x) {x^n\over n!} = \ha\int dx f''(x){\hat D}_{N-1}(x){x^n\over n!}
= (-)^{n-1}T_{N-n}
\\
n=0,1,2,\ldots,\ \ \ \ T_0=-a
\ee
which have to be taken into account, if one considers variation of the functional over the
new density.

In other words, instead of \rf{functnl} one can consider an extremum for
\be
\label{fuN}
\F = \F_N\left[\rho\right] = {1\over 2}\int dx \rho(x){\bf t}_N(x) -
{(-)^{N-1}\over 2(2N)!}\int_{x_1>x_2} dx_1 dx_2 \rho(x_1)\rho(x_2) H_{2N}^{(+)}(x_1-x_2) +
\\
+\sum_{n=0}^N\sigma_n\left(T_n-{(-)^{n-1}\over 2}\int dx {x^{N-n}\over (N-n)!}\rho(x)\right)
\ee
where the kernel
${(-)^{N-1}\over (2N)!}H_{2N}^{(+)}(x) = {1\over (2N)!}x^{2N}\left( \log x - c_{2N}\right)$
does not depend explicitly of the times ${\bf T}$, all this dependence is absorbed by
density $\rho(x)$.
The extremum condition for the functional \rf{fuN} stays, that (real part of)
\be
\label{exN}
S_{N}(z) = {\bf t}_N(z) - {(-)^{N-1}\over (2N)!}\int dx \rho(x) H_{2N}^{(+)}(z-x) +
\sum_{n=0}^N \sigma_n (-)^{n}{z^{N-n}\over (N-n)!}
\ee
vanishes on the support $z\in{\bf I}$ of $\rho(z)$. Taking up to $N$-th derivatives of \rf{exN} one
gets
\be
\label{derexN}
S_{N-1}(z) = \textbf{t}_{N-1}(z) - {(-)^{N-1}\over (2N-1)!}\int dx \rho(x) H_{2N-1}^{(+)}(z-x) +
\sum_{n=0}^{N-1} \sigma_n (-)^{n}{z^{N-n-1}\over (N-n-1)!}
\\
\vdots
\\
S(z) =\textbf{ t}'(z) - {(-)^{N-1}\over N!}\int dx \rho(x) H_{N}^{(+)}(z-x) + \sigma_0
\ee
a sequence of functions vanishing on the cut. The last integral $S(z) ={d^{N}\over dz^{N}}S_N(z)$
coincides with \rf{Ssigma}, and therefore has the same properties.

In particular, at $z\to\infty$ the last integral in \rf{derexN} has an expansion
where the coefficients are expressed by the ``moments'' of new density
\be
\label{SNexp}
S(z)\ \stackreb{z\to\infty}{=}\ {\bf t}'(z) -
\sum_{n=0}^N z^n(\log z-c_n) {(-)^{N-n}\over (N-n)!}\int x^{N-n}\rho (x)dx + \sigma_0 +
\\
- \sum_{k>0} {1\over k(k+1)\ldots(k+N)z^k}\int x^{N+k}\rho(x) dx =
\\
= \textbf{t}'(z) - 2\sum_{n=0}^N T_n z^n(\log z-c_n) - {\d\F\over \d a}
- 2\sum_{k>0} {1\over kz^k}{\d\F\over\d t_k}
\ee
reproducing \rf{sasymE} by
\rf{rhof} and \rf{rhocstr} (or upon the constraints at Lagrange multipliers in \rf{fuN}).
From the properties of the functional \rf{fuN}, one can
straightforwardly find the derivatives
\be
\label{derFN}
{\d\F\over\d t_k} = {1\over 2(k+1)\ldots(k+N)}\int x^{k+N}\rho(x) dx
\\
\sigma_0 = {\d\F\over\d T_0} = - {\d\F\over\d a} = - a^D
\ee
coinciding with \rf{derfunt}.

However, after arbsorbing all nontrivial ${\bf T}$-dependence into $\rho$ in \rf{fuN},
it becomes obvious, that
\be
\label{sigmasem}
{\d\F\over\d T_n} = \sigma_n = (-)^{n} n!\left(S_n\right)_0, \ \ \ n=0,\ldots,N
\ee
The naively divergent integrals, containing $\rho(x)$, should be
understood only in the sense of \rf{rhof}.

\setcounter{equation}0
\section{Nonabelian theory from abelian integrals
\label{sect:nonabel}}

Finally, let us turn to discussion of generic nonabelian theory, whose perturbative limit
was considered in sect.~\ref{sect:pert}. The quasiclassical tau-function is now defined
by constructing an abelian integral on the hyperelliptic curve \rf{dcN}, whose properties
can be extracted from integral representations of sect.~\ref{sect:func}.

It is again important to fix certain finite number $N$ of the gravitational descendants of unity being
switched on. The integral representation \rf{Ssigma} defines a multivalued abelian integral on
the curve \rf{dcN}, and only its $N$-th derivative becomes single-valued. Denote as usual
$\Phi = {dS\over dz}$, and further $\Phi'={d\Phi\over dz},\ldots$ up to
\be
\label{fin1}
d\Phi^{(N-1)} = d\left({d^{N-1}\Phi\over dz^{N-1}}\right)=
\\
= {\bf t}^{(N+2)}(z)dz
- N!T_N\int {f''(x)dx\over z-x} dz
- \sum_{k=1}^{N-1}(-)^k \int {T^{(N-k)}(x)f''(x)dx\over (z-x)^{k+1}} dz
\ee
which is already a single-valued on the non-degenerate curve \rf{dcN} abelian differential,
odd under the hyperelliptic
involution, since its real part vanishes on the cut. Its form can be totally determined
by its singularities at the infinity points $P_\pm$ and at the ramification points $\{ x_j\}$,
$j=1,\ldots,2N_c$, where it also has poles due to behavior $f''(x) \sim (x-x_\pm)^{-1/2}$.
The singularities at ramification points are in fact artificial, in the sense that one may
think of $\Phi',\ldots,\Phi^{(N-1)}$ as of the regular at branch points
$2-,\ldots,N-$ differentials on the curve \rf{dcN}.

One can therefore write for \rf{fin1} an explicit formula
\be
\label{fin1hy}
d\Phi^{(N-1)} = {\phi(z)dz\over y} + {dz\over y}\sum_{j=1}^{2N_c}\sum_{k=1}^{N-1}
\left({q^k_j\over(z-x_j)^k}\right)
\ee
where $\phi(z)$ is a polynomial of power
\be
\label{degfi}
\deg\phi(z) = \left\{
\begin{array}{c}
  N_c-1,\ \ \ n\leq N \\
  N_c-1+n-N,\ \ \ n>N
\end{array}\right.
\ee
for the theory on genus $N_c-1$ curve \rf{dcN}
and with $n-1$ and $N$ nonvanishing times $\{ t_k\}$ and $\{ T_n\}$ correspondingly.
The periods of \rf{fin1hy} are fixed by \rf{fin1}, or
\be
\label{perfin1}
{1\over 2\pi i}\oint_{A_i} d\Phi^{(N-1)} = - 2N! T_N
\\
\oint_{B_i} d\Phi^{(N-1)} = 0
\ee
Couning the period constraints \rf{perfin1},
one can consider $N_c$ cycles $A_k$, $k=1,\ldots,N_c$, surrounding
generally $N_c$ distinct segments of the support of $f''(x)\neq 0$, $x\in {\bf I}_k,\ k=1,\ldots,N_c$,
which is equivalent to the canonical choice of $A$-cycles together with the residue at infinity.
Totally, \rf{perfin1} give $2N_c-1$ period constraints, and should be completed by the $2N_c$-th
condition
\be
\label{torus}
\int_{z(P_-)}^{z(P_+)} d\Phi^{(N-1)} = -2N!N_cT_N\log z + 4\pi i N!T_N\mathbb{Z} +
O\left({1\over z}\right)
\ee
i.e. the regularized constant part of the integral $\int_{P_-}^{P_+} d\Phi^{(N-1)}$ vanishes
modulo the period lattice \rf{perfin1}, since the integral \rf{torus} depends on the choice of
the integration path.

\bigskip
\mezzo{Small phase space and $T_2\neq 0$}

\noindent
Consider for simplicity only $t_1\neq 0$ and switched on $T_1,T_2$. Formula \rf{fin1} gives
for this case
\be
\label{dFip}
d\Phi' = {\phi_{N_c-1}(z)dz\over y} + {dz\over y}\sum_{j=1}^{2N_c}
\left({q_j\over z-x_j}\right)
\ee
which depends on $3N_c$ coefficients of $\phi_{N_c-1}(z)$ and $\{ q_j\}$, as well as
$2N_c$ branch points $\{ x_j\}$, i.e. totally of $5N_c$ undetermined yet coefficients.
The period integrals \rf{perfin1}, together with the residue
\be
\label{resdFip}
\res_{P_+}d\Phi' = - \res_{P_-}d\Phi' = 4T_2\cdot N_c
\ee
give altogether $2N_c$ constraints, or fix the parameters $\{ q_j\}$ of the differential \rf{dFip},
leaving yet no restrictions for the coefficients of $\phi_{N_c-1}(z)$ and branch points
of the curve.

Now, one can define an abelian integral $\Phi'(P) = \int^P d\Phi'$ or the differential
\be
\label{dFi}
d\Phi = dz\int^z d\Phi'
\ee
which is multivalued, but all the jumps are fixed by \rf{perfin1}, being proportional to
$4\pi i\cdot T_2 dz$. The integration constant in \rf{dFi} is fixed by requirement, that
$\Phi'(P)\ \stackreb{z(P)\to\infty}{\sim}\ -4N_cT_2\log z + O\left({1\over z}\right)$,
consistent due to \rf{torus}.
Since the differential of hyperelliptic co-ordinate on \rf{dcN}
has vanishing periods $\oint dz=0$ along any cycle, one can make sense of the periods
of the differential \rf{dFi} itself, and put
\be
\label{perdFi}
{1\over 2\pi i}\oint_{A_k} d\Phi \equiv
- {1\over 2\pi i}\oint_{A_k} zd\Phi' = T_1\int_{{\bf I}_k} dxf''(x) = 2T_1,\ \ \ k=1,\ldots,N_c
\\
\oint_{B_k} d\Phi \equiv \oint_{B_k} zd\Phi' = 0
\ee
The period integrals \rf{perdFi} together with normalization condition (say, $\Phi(x_{N_c})=0$)
give $2N_c$ more constraints on the total set of undetermined parameters, while the rest
is absorbed by the Seiberg-Witten periods, defined now as
\be
\label{SWAN2}
a_j = {1\over 4\pi i}\oint_{A_j} {z^2\over 2}d\Phi',\ \ \ \ j=1,\ldots,N_c
\ee
whose sum gives the residue at infinity.

\bigskip
\mezzo{$N$ descendants $T_1,\ldots,T_N\neq 0$}

\noindent
Almost the same counting can be performed for the generic case with $N$ descendants.
One has now $2N_c\cdot N+N_c=(2N+1)N_c$ parameters of $\phi_{N_c-1}(z)$, $\{ q^k_j\}$ and
branch points $\{ x_j\}$ (in the case of nonvanishing higher $\{ t_k\}$ they will be
absorbed into higher coefficients of the polynomial $\phi(z)$ and the integration constants).
Being constrained by constancy of its periods, we rest with $(2N-1)N_c$ variables.

We have then to restore the differential $d\Phi$ by multiple integration of \rf{fin1}. At each
step we have to fix the periods of $d\Phi^{(N-2)},\ldots,d\Phi'$ by $2N_c$ constraints,
ending up, therefore with
\be
\label{count}
(2N+1)N_c - 2N_c\cdot N = N_c
\ee
variables, which can be conveniently chosen as the Seiberg-Witten periods
\be
\label{SWperA}
a_j = {1\over 4\pi i}\oint_{A_j} {z^N\over N!}d\Phi^{(N-1)},\ \ \ \ j=1,\ldots,N_c
\ee
The multivalued differential $dS = \Phi dz$ has now constant jumps, depending linearly
upon $a$ and the times $T_1,\ldots,T_N$, and one can always choose its branch with the
asymptotic \rf{sasymE}, if taken along the real axis at $z\to +\infty$ on the
``upper'' sheet.

\bigskip
\mezzo{Quasiclassical tau-function}

\noindent
The dual periods
\be
\label{SWperB}
a^D_j = \half\oint_{B_j} {z^N\over N!}d\Phi^{(N-1)} = {\d\F\over\d a_j},\ \ \ \ j=1,\ldots,N_c
\ee
define the gradients of the quasiclassical tau-function. Integrability condition for
\rf{SWperB} is guaranteed by symmetricity of the period matrix of the curve \rf{dcN}, following
from
\be
\label{delShol}
\delta (dS) = \delta \left(\Phi dz\right)
\simeq {\rm holomorphic}
\ee
following directly from the constancy of the periods $d\Phi',\ldots,d\Phi^{(N-1)}$. In addition
to the remaining intact
``abelian formulas'' \rf{tPdT} and \rf{TdF} that defines the full quasiclassical tau-function for the
perturbed theory, and the integrability is guaranteed by the Riemann bilinear relations.

\setcounter{equation}0
\section{Discussion
\label{sect:disc}}

We have presented in this paper a quasiclassical geometric formulation for the full non-truncated
topological $\mathbb{P}^1$ string model, when all the descendants $\sigma_k(\varpi)$ and
$\sigma_k({\bf 1})$ with $k>0$ are switched on, and propose its generalization to the nonabelian
dual supersymmetric gauge theory. For the topological string model the quasiclassical formulation
is given in ``mirror'' terms - a rational curve, which can be interpreted as a dual Landau-Ginzburg
superpotential $z=v+\Lambda\left(w+{1\over w}\right)$ on a cylinder, and the set of functions, odd
under its involution $w\leftrightarrow{1\over w}$. The descendants of the K\"ahler class
$\sigma_k(\varpi)$ generate the flows of dispersionless Toda chain hierarchy, while the descendants
of unity $\sigma_k({\bf 1})$ produce the logarithmic flows \cite{EY} of the so called \cite{DubTod}
extended Toda hierarchy, which can be possibly reformulated as a reduction of two-dimensional
Toda lattice. The exact relation of the quasiclassical solution, proposed above, to the two-dimensional
Toda lattice is beyond the scope of this paper, but let us present here a hint, how the multiple
integral formula \rf{TdF} can be interpreted in this way.

\bigskip
\mezzo{Equivariant Toda lattice}

\noindent
The relation between the extended Toda and equivariant Toda lattice \cite{op,Milan} includes
the change of the variables
\be
\label{xtT}
X_{k+1} = {T_k\over\epsilon} + t_{k+1},\ \ \ \
{\bar X}_{k+1} = - {T_k\over\epsilon},\ \ \ \ k\geq 0
\ee
or
\be
\label{tTx}
t_k = X_k + {\bar X}_k,\ \ \ \ k>0
\\
T_k = -\epsilon{\bar X}_{k+1},\ \ \ \ k\geq 0
\ee
For example, the prepotential on the small phase space
\be
\label{eFsm}
\F (X_1,{\bar X}_1;\epsilon) = {\epsilon^2\over 6}\left(X_1^3+{\bar X}_1^3\right) + e^{X_1+{\bar X}_1} =
\\
= {a^2t_1\over 2}+e^t_1+{\epsilon\over 2}at_1^2+{\epsilon^2\over 6}t_1^3
\ee
coinciding with \rf{Fsphs} at $\epsilon\to 0$,
indeed satisfies the two-dimensional Toda lattice equation
\be
\label{eToda}
{\d^2\F\over\d{\bar X}_1\d X_1} =
\exp\left({1\over\epsilon^2}\left({\d\over\d X_1}-{\d\over\d {\bar X}_1}\right)^2\F\right)
\ee
if one takes the solutions, constrained by reduction, including
\be
\label{red2Toda1}
{\d\F\over\d X_1}-{\d\F\over\d {\bar X}_1} = \epsilon{\d\F\over\d X_0}
\ee
One can expect therefore, generally, that
\be
\label{red2Toda}
{\d\F\over\d X_k}-{\d\F\over\d {\bar X}_k} = \epsilon {\hat R}_k\circ\F,\ \ \ \forall k>0
\ee
where ${\hat R}_k$ is presumably a ($k$-th order) differential operator in $X_0$,
${\hat R}_1 = {\d\over\d X_0}$. At $\epsilon\to 0$ conditions \rf{red2Toda}, \rf{red2Toda1}
turn into the Toda chain reduction
\be
\label{redTodach}
{\d\F\over\d X_k}-{\d\F\over\d {\bar X}_k} = 0,\ \ \ \forall k>0
\\
{\d\F\over\d X_k}+{\d\F\over\d {\bar X}_k} =2{\d\F\over\d t_k},\ \ \ \forall k>0
\ee
where $\{ t_k\}$ \rf{tTx} are the times of the Toda chain. More generally, in the
reduction to the Toda chain, the first set of conditions \rf{redTodach} can have a linear
function at the r.h.s.
\be
\label{redTodali}
{\d\F\over\d X_k}-{\d\F\over\d {\bar X}_k} = C_k(X_k-{\bar X}_k),\ \ \ \forall k>0
\ee
with $C_k\sim k$ is a constant as a function of times. For the function \rf{eFsm} one gets
instead if \rf{redTodali}
\be
{\d\F\over\d X_1}-{\d\F\over\d {\bar X}_1} = {\epsilon\over 2}\left(X_1^2-{\bar X}_1^2\right) =
{\epsilon\over 2}\left(X_1+{\bar X}_1\right)\left(X_1-{\bar X}_1\right)
\ee
so one finds, that $C_1={\epsilon\over 2}\left(X_1+{\bar X}_1\right)={\epsilon\over 2}t_1$
instead of a constant becomes a ``slow'' modulated linear function of the Toda chain time $t_1$.
The exact form of the operators ${\hat R}_k$ is not yet known (though perhaps can be extracted
from \cite{op}), but the formulas \rf{TdF}, \rf{TdFint} establish the quasiclassical correspondence
\be
\label{qqcorr}
{\hat R}_n \F\ \sim\ \left[\underbrace{\int_{dz}\ldots\int_{dz}}_n\ S\right]_0
\ee
For the two-dimensional Toda lattice one has two different co-ordinates $z_+$ and $z_-$ at
two infinities $P_\pm$ corresponding to the flows in $X$ and ${\bar X}$ time variables. One
may think then, that $z_+ - z_- \sim \int dS$ and the differences of the higher Hamiltonians
$\Omega(z_+)-{\tilde\Omega}(z_-) \sim {\int\ldots\int}\ dS$ produces
the desired formula \rf{qqcorr}.

\bigskip
\mezzo{Nonabelian theory}

\noindent
It is not yet completely clear, when is the sense of ``descendant'' deformation of the nonabelian
theory. The descendants of the K\"ahler class deform the gauge theory in the ultraviolet, which
is encoded in $\half\tau_0 x^2 \rightarrow F_{UV}(x;{\bf t})$ for the short-distance prepotential
\rf{fuv}. The descendants of unity perform rather a reparameterization on the moduli space of
gauge theory $a_j \rightarrow T(a_j) + O(\Lambda^{2N_c})$, whose exact sense remains yet unclear.

We have considered in \cite{MN,AMpopya} and above here the theory, where all $t_k$ with $k>1$ and
$T_n$ with $n>1$ generate infinitesimal perturbations of the model on ``small phase space". Nevertheless,
all descendants deform the Seiberg-Witten curve (except for the ``abelian'' case of the
$\mathbb{P}^1$ model), which now turns to be a generic hyperelliptic curve \rf{dcN}, though still
being ``not to far'' in the moduli space from the Seiberg-Witten curve \rf{Todacu}. In particular,
we do not address any questions, related with possible ``large'' deformations in moduli space,
changing the genus etc. Roughly speaking, if the ${\bf t}$-deformations of the theory lead us
towards the processes of generation of fundamental multiples, in the same sense the $T$-deformations
lead towards embedding of the theory into the compactified higher-dimensional target spaces.

\bigskip
\mezzo{Acknowledgements}

\noindent
I am grateful to A.~Alexandrov, M.~Kazarian, S.~Kharchev, I.~Krichever, A.~S.~Losev, A.~Mironov,
and N.~Nekrasov for the very useful discussions. I would like to thank Institut des Hautes
Etudes Scientifiques at Bures-sur-Yvette and Laboratoire de Mathematiques et Physique
Theorique at Universite Francois Rabelais, Tours, where essential parts of this work
have been done, and the Max Planck Institute for Mathematics in Bonn, where it has been completed.

The work was partially supported by the Federal Nuclear Energy Agency, the RFBR grant
06-02-17383,
the grant for support of Scientific Schools 4401.2006.2,
INTAS grant 05-1000008-7865, the project ANR-05-BLAN-0029-01, the
NWO-RFBR program 047.017.2004.015, the Russian-Italian RFBR program 06-01-92059-CE, and by the
Dynasty foundation.

\section*{Appendix}
\appendix
\setcounter{equation}0
\section{Riemann bilinear identities
\label{ap:RBI}}

Equation \rf{TdF}, or
\be
\label{TdFint}
\left.{\d\F\over \d T_n}\right|_{\bf t} = \left[\underbrace{\int_{dz}\ldots\int_{dz}}_n\ S\right]_0
\ee
gives rise to the mixed second derivatives
\be
\label{TtdF}
{\d^2\F\over \d t_k\d T_n} = \left[\underbrace{\int_{dz}\ldots\int_{dz}}_n\ \Omega_k\right]_0
\ee
which should be compared to
\be
\label{tTdF}
{\d^2\F\over \d T_n\d t_k} = \ha \res_{P_+} z^{k}dH_n
= - \ha \res_{P_-} z^{k}dH_n,
\ee
following from \rf{tPd}. In order to establish equivalence between \rf{TtdF} and \rf{tTdF},
consider the integral along the boundary of the cut $w$-cylinder with the removed points
$P_\pm$
\be
\label{intbou}
\oint_{\d\Sigma}H_n d\Omega_k = 2\pi i\sum \res\ H_n d\Omega_k = 0
\ee
The integral in the l.h.s. can be rewritten as
\be
\label{intbouAB}
\oint_{\d\Sigma}H_n d\Omega_k = \left[\int_{A^+}+\int_{A^-}\right]H_n d\Omega_k
+ \left[\int_{B^+}+\int_{B^-}\right]H_n d\Omega_k
\ee
where we have chosen the following parameterization of the cut $w$-cylinder:
\be
\label{wcyl}
A^+:\ \ \ w=\varepsilon e^{i\varphi},\ \ \ 0<\varphi<2\pi
\\
B^+:\ \ \ \varepsilon<w<R
\\
A^-:\ \ \ w=R e^{i\varphi},\ \ \ 2\pi>\varphi>0
\\
B^-:\ \ \ R>w>\varepsilon
\ee
The last term in the r.h.s. of \rf{intbouAB} gives
\be
\label{intB}
\left[\int_{B^+}+\int_{B^-}\right]H_n d\Omega_k =
\int_B \left(H_n^+d\Omega_k - H_n^-d\Omega_k\right) = 2\pi i\int_B z^n d\Omega_k =
\\
= 2\pi i\left.\left( z^n\Omega_k - nz^{n-1}\Omega_k^{(1)} + \ldots + (-)^n n!\Omega_k^{(n)}
\right)\right|^R_\varepsilon
\ee
where, similarly to \rf{Sn},
\be
\label{On}
{d^n \Omega_k^{(n)}\over dz^n} = \Omega_k,\ \ \ \ n\geq 0
\ee
is introduced.

For the $A$-integrals one can write
\be
\label{intA}
\int_{A^\pm}H_n d\Omega_k = \int_{A^\pm}H_n^{(+)}(z)d\Omega_k\ \mp \
2\pi i\ \res_{P_\pm} z^k dH_n
\ee
where by residue the coefficient at the term $z^{-1}$ is meant. The first term
in the r.h.s. of \rf{intA} can be integrated by parts using \rf{HEas} and \rf{On},
giving rise to
\be
\label{iAbp}
\int_{A^+}H_n^{(+)} d\Omega_k = \left.\left( H_n^{(+)}\Omega_k - nH_{n-1}^{(+)}\Omega_k^{(1)} +
\ldots + (-)^n n!H_0^{(+)}\Omega_k^{(n)}\right)\right|^{\varepsilon^+}_{\varepsilon^-} =
\\
= 2\pi i\left.\left( z^n\Omega_k - nz^{n-1}\Omega_k^{(1)} + \ldots + (-)^n n!\Omega_k^{(n)}
\right)(\varepsilon)\right|_{\rm div}
\\
\int_{A^-}H_n^{(+)} d\Omega_k = \left.\left( H_n^{(+)}\Omega_k - nH_{n-1}^{(+)}\Omega_k^{(1)} +
\ldots + (-)^n n!H_0^{(+)}\Omega_k^{(n)}\right)\right|^{R^-}_{R^+}=
\\
= - 2\pi i\left.\left( z^n\Omega_k - nz^{n-1}\Omega_k^{(1)} + \ldots + (-)^n n!\Omega_k^{(n)}
\right)(R)\right|_{\rm div}
\ee
where in the r.h.s.'s one gets only the {\em divergent} parts of the corresponding
expressions.

Altogether \rf{intbou}, \rf{intbouAB}, \rf{intB}, \rf{intA} and \rf{iAbp} give rise to
\be
\label{intres}
0 = \left.\left( z^n\Omega_k - nz^{n-1}\Omega_k^{(1)} + \ldots + (-)^n n!\Omega_k^{(n)}
\right)\right|^R_\varepsilon +
\\
+ \left.\left( z^n\Omega_k - nz^{n-1}\Omega_k^{(1)} + \ldots + (-)^n n!\Omega_k^{(n)}
\right)(\varepsilon)\right|_{\rm div} -
\\
- \left.\left( z^n\Omega_k - nz^{n-1}\Omega_k^{(1)} + \ldots + (-)^n n!\Omega_k^{(n)}
\right)(R)\right|_{\rm div} -
\\
- \res_{P_+} z^k dH_n + \res_{P_-} z^k dH_n
\ee
or, using the antisymmetry w.r.t. involution exchanging $P_+$ and $P_-$,
\be
\label{rbires}
\res_{P_+} z^k dH_n = - \res_{P_-} z^k dH_n =
\\
= \left.\left( z^n\Omega_k - nz^{n-1}\Omega_k^{(1)} + \ldots + (-)^n n!\Omega_k^{(n)}
\right)(\varepsilon)\right|_{\rm const} =
\\
= -
\left.\left( z^n\Omega_k - nz^{n-1}\Omega_k^{(1)} + \ldots + (-)^n n!\Omega_k^{(n)}
\right)(R)\right|_{\rm const}
\ee
or
\be
\label{rbi}
\res_{P_+} z^k dH_n = - \res_{P_-} z^k dH_n = (-)^n n! \left[\Omega_k^{(n)}\right]_0
\ee


\begin{thebibliography}{99}

\bibitem{LMN}
  A.~S.~Losev, A.~Marshakov and N.~Nekrasov,
  ``Small instantons, little strings and free fermions,'' in Ian Kogan memorial volume, M.Shifman,
A.Vainshtein and J. Wheater (eds.) {\em From fields to strings:
circumnavigating theoretical physics}, 581-621
  [arXiv:hep-th/0302191].


\bibitem{MN}
A.~Marshakov and N.~Nekrasov,
  JHEP {\bf 0701} (2007) 104
  [arXiv:hep-th/0612019].

\bibitem{Nek}
  N.~Nekrasov,
  Adv.\ Theor.\ Math.\ Phys.\  {\bf 7} (2004) 831
  [arXiv:hep-th/0206161].


\bibitem{KriW}
I.~Krichever,
Commun. Pure. Appl. Math. {\bf 47} (1992) 437
[arXiv:hep-th/9205110].


\bibitem{EY}
  T.~Eguchi and S.~K.~Yang,
  Mod.\ Phys.\ Lett.\  A {\bf 9} (1994) 2893
  [arXiv:hep-th/9407134].

\bibitem{EHY}
  T.~Eguchi, K.~Hori and S.~K.~Yang,
  Int.\ J.\ Mod.\ Phys.\  A {\bf 10} (1995) 4203
  [arXiv:hep-th/9503017].

\bibitem{giv}
A.~Givental, ``Gromov-Witten invariants and quantization of quadratic hamiltonians'',
arXiv:math/0108100.

\bibitem{op}
A.~Okounkov and R.~Pandharipande, ``Gromov-Witten theory, Hurwitz theory, and completed cycles'',
arXiv:math.AG/0204305;
``The equivariant Gromov-Witten theory of $P^1$'', arXiv:math.AG/0207233.

\bibitem{opvir}
A.~Okounkov and R.~Pandharipande, ``Virasoro constraints for target curves'',
arXiv:math/0308097.

\bibitem{DubTod}
G.~Carlet, B.~Dubrovin and Y.~Zhang, ``The extended Toda hierarchy'',
arXiv:nlin/0306060;\\
B.~Dubrovin and Y.~Zhang, ``Virasoro symmetries of the extended Toda hierarchy'',
arXiv:math/0308152.

\bibitem{Milan}
T.~Milanov, ``Hirota quadratic equations for the extended Toda hierarchy'',
arXiv:math/0501336; ``The equivariant Gromov-Witten theory of $CP^1$ and integrable hierarchies'',
arXiv:math-ph/0508054; ``Gromov-Witten theory of $CP^1$ and integrable hierarchies'',
arXiv:math-ph/0605001.

\bibitem{AMpopya}
  A.~Marshakov,
  Theor. Math. Phys. {\bf 154} (2008) 362-384
(Teor. Mat. Fiz. {\bf 154} (2008) 424-450); arXiv:0706.2851 [hep-th].

\bibitem{NO}
  N.~Nekrasov and A.~Okounkov,
  ``Seiberg-Witten theory and random partitions,''
  arXiv:hep-th/0306238.

\bibitem{Lossev}
K.~Saito, ``On the periods of primitive integrals", Harvard Lecture Notes, 1980;\\
A.~S.~Losev,
  Theor.\ Math.\ Phys.\  {\bf 95} (1993) 595
  [Teor.\ Mat.\ Fiz.\  {\bf 95} (1993) 307]
  [arXiv:hep-th/9211090];\\
T.~Eguchi, H.~Kanno, Y.~Yamada and S.~K.~Yang,
  Phys.\ Lett.\  B {\bf 305} (1993) 235
  [arXiv:hep-th/9302048].

\bibitem{VK}
B.~Logan and L.~Shepp,
Advances in Math. 26 (1977), no. 2, 206;\\
S.~Kerov and A.~Vershik,
DAN SSSR, {\bf 233},1024(1977), (in Russian).

\bibitem{5dSW}
N.~Nekrasov,
  Nucl.\ Phys.\   {\bf B531} (1998) 323
  [arXiv:hep-th/9609219];\\
A.~Lawrence and N.~Nekrasov,
  Nucl.\ Phys.\   {\bf B513} (1998) 239
  [arXiv:hep-th/9706025];\\
A.~Marshakov and A.~Mironov,
  Nucl.\ Phys.\  B {\bf 518} (1998) 59
  [arXiv:hep-th/9711156];\\
H.~Braden, A.~Marshakov, A.~Mironov and A.~Morozov,
  Phys.\ Lett.\  B {\bf 448} (1999) 195
  [arXiv:hep-th/9812078];
  Nucl.\ Phys.\  B {\bf 558} (1999) 371
  [arXiv:hep-th/9902205].


\bibitem{5dTak}
T.~Maeda, T.~Nakatsu, K.~Takasaki and T.~Tamakoshi,
  Nucl.\ Phys.\  B {\bf 715} (2005) 275
  [arXiv:hep-th/0412329];\\
T.~Nakatsu and K.~Takasaki,
  ``Melting crystal, quantum torus and Toda hierarchy,''
  arXiv:0710.5339 [hep-th].

\end{thebibliography}
\end{document}